\documentclass[twocolumn,aps,prd,superscriptaddress,nofootinbib,floatfix,preprintnumbers]{revtex4-1}

\usepackage{graphicx,multirow}
\usepackage{xspace}

\usepackage{amsmath,amsfonts}

\usepackage{hyperref}
\usepackage{soul}

\newcommand{\beq}{\begin{equation}}
\newcommand{\eeq}{\end{equation}}
\newcommand{\beqa}{\begin{eqnarray}}
\newcommand{\eeqa}{\end{eqnarray}}

\newcommand{\babar}{\mbox{\ensuremath{{\displaystyle B}\!{\scriptstyle A}{\displaystyle B}\!{\scriptstyle AR}}}\xspace}

\newcommand{\Bbar}{\,\overline{\!B}{}}
\newcommand{\Dbar}{\,\overline{\!D}{}}
\newcommand{\Kbar}{\,\overline{\!K}{}}
\def\B0bar{\Bbar{}^0}
\def\D0bar{\Dbar{}^0}
\def\K0bar{\Kbar{}^0}

\begin{document}

\preprint{arXiv:1509.06938}

\title{\boldmath The $B \to \pi \, \tau \, \bar \nu_\tau$ decay in the context of the 2HDM type II}

\author{Florian U.\ Bernlochner}
\affiliation{Physikalisches Institut der Rheinische Friedrich-Wilhelms-Universit\"at Bonn, 53115 Bonn, Germany}

\begin{abstract}
In this manuscript the $B \to \pi \, \tau \, \bar \nu_\tau$ decay is investigated in the context of the 2HDM type II extension of the SM. In particular, a prediction for the ratio of semileptonic branching fractions from $\ell = \tau$ and $\ell = e , \mu$ is produced and an exclusion based on the MSSM parameters $\tan \beta$ and $m_{H}^+$ from a recent Belle measurement is reported. An alternative variable to the ratio of branching fractions that could offer increased sensitivity is discussed that could result in increased sensitivity to rule out contributions from additional scalar mediators. 
\end{abstract}

\maketitle

\section{Introduction}\label{sec:intro}

In recent measurements the $B$-Factory experiments \babar, Belle and LHCb have reported large disagreements in semi-tauonic decays involving ratios of charmed final states~\cite{Lees:2012xj,Lees:2013uzd,Huschle:2015rga,Aaij:2015yra}
. A recent average from Ref.~\cite{HFAG} quantifies the overall disagreement with the Standard Model (SM) expectation at the 3.9 $\sigma$ level. This disagreement is an interesting anomaly that deserves more experimental and theoretical study, in particular due to the absence of any clear new physics signal from the experiments at the energy frontier of particle physics. An interesting alternative probe to see similar effects is given with $B \to \pi \, \tau \, \bar \nu_\tau$: although CKM suppressed, the $B \to \pi \, \tau \, \bar \nu_\tau$ decay is less phase space suppressed with respect to the transition involving light leptons $\ell = e, \mu$ as its charmed partner decays, resulting in a high fraction of $B \to \pi \, \tau \, \bar \nu_\tau$ over $B \to \pi \, \ell \, \bar \nu_\ell$ in the SM. In this manuscript a prediction for the ratio 
\begin{align} \label{eq:rpi}
 R_\pi & = \frac{ \mathcal{B}(B \to \pi \, \tau \, \bar \nu_\tau)}{\mathcal{B}(B \to \pi \, \ell \, \bar \nu_\ell)}  \, ,
\end{align}
for the SM is given and its modifications in the context of the two-Higgs doublet model (2HDM) of type II are discussed. A first exclusion of the MSSM parameters $\tan \beta$ and $m_{H^+}$ is carried out using recent limits on the $B \to \pi \, \tau \, \bar \nu_\tau$ branching fraction reported by the Belle experiment~\cite{bellepitaunu}. As will be shown, Equation~\ref{eq:rpi} can be predicted with a precision of a few  percent as many of the theoretical uncertainties associated with the hadronic form factors cancel and no dependence on the CKM matrix element $\left| V_{ub} \right|$ remains. Similar predictions have been discussed in Refs.~\cite{Tanaka:1994ay,Chen:2006nua,Kim:2007uq,Khodjamirian:2011ub,Dutta:2013qaa}. Recent progress in the understanding of the $B \to \pi$ form factor as well as the availability of first experimental results give a strong motivation to revisit this topic. 

This manuscript is organized as follows: Section~\ref{sec:smdecay} briefly reviews the $B \to \pi \, \tau \, \bar \nu_\tau$ decay in the SM and summarizes the state-of-the-art knowledge of the $B \to \pi$ form factor. Section~\ref{sec:mssm} discusses the modifications of the decay rate in the context of the two Higgs doublet model (2HDM) type II interactions. In Section~\ref{sec:rpi} the prediction of the ratio of branching fractions from $B \to \pi \, \tau \, \bar \nu_\tau$  and $B \to \pi \, \ell \, \bar \nu_\ell$ is discussed and the exclusion of the 2HDM type II parameter space is presented. Section~\ref{sec:rpiq2} discusses a more sensitive variable relying on the reconstruction of the four-momentum transfer squared and the manuscript concludes in Section~\ref{sec:concl} with a summary of the key results. 

\section{The $B \to \pi \, \tau \, \bar \nu_\tau$ decay in the SM}\label{sec:smdecay}

The effective SM Lagrangian describing the $b \to u \, \ell \, \bar \nu_\ell$\footnote{In this section $\ell$ denotes any massive lepton unless stated otherwise.} transition is well known in the literature and given by 
\begin{align}
\mathcal{L}_{\rm eff} =  \frac{- 4 G_F} { \sqrt{2}} \, V_{ub} \, \left( \bar u \gamma_\mu P_L b  \right) \left( \bar \nu \gamma^\mu P_L \ell  \right) + \text{h.c.}
\end{align}
with the projection operator $P_L = \left( 1 - \gamma_5 \right) / 2$ and $G_F$ Fermi's constant. The $B \to \pi \, \ell \, \bar \nu_\ell$ decay amplitude depends on one non-perturbative hadronic matrix element that can be expressed using Lorentz invariance and the equation of motion in terms of $B \to \pi$ form factors. The $B \to \pi$ transition form factors $f_{+/0}$ are defined by 
\begin{align} \label{eq:hadcurrent}
 \langle \pi (p_\pi) | \bar u \, \gamma_\mu \, P_L \, b | B(p) \rangle = &  f_+(q^2) \left( (p + p_\pi)_\mu - \frac{m_B^2 - m_\pi^2}{q^2} \, q^\mu  \right) \nonumber \\ 
 &  + f_0(q^2) \frac{ m_B^2 - m_\pi^2 }{q^2} \, q^\mu \, ,
\end{align}
with $q = p - p_\pi$ denoting the four-momentum transfer between the $B$-meson and the final-state pion of the semileptonic decay. Further, $m_B$ denotes the $B$-meson mass and $m_\pi$ the pion mass. The differential decay rate as a function of $q^2$ with its full lepton mass dependence is given by 
\begin{align} \label{eq:rate}
  \frac{\text{d} \Gamma(B \to \pi \, \ell \, \bar \nu_\ell) }{\text{d} q^2} & = \frac{8 \left| \vec p_\pi \right|  }{3} \frac{ G_F^2 \, \left| V_{ub} \right|^2 q^2 }{256 \, \pi^3 \, m_B^2} \, \left( 1 - \frac{m_\ell^2}{q^2} \right)^2 \nonumber \\
    & \bigg[  H_0^2(q^2) \left( 1 + \frac{m_\ell^2}{2 q^2} \right) + \frac{3}{2} \frac{m_\ell^2}{q^2} \, H_t^2(q^2) \bigg] \, ,
\end{align}
with the Helicity amplitudes $H_0$ and $H_t$ and  $\left| \vec p_\pi \right|$ the absolute three-momentum of the final state pion. The absolute three-momentum is related to the four-momentum transfer squared as
\begin{align}
\left| \vec p_\pi \right| & = \sqrt{ \left( \frac{m_B^2 + m_\pi^2 - q^2}{2 m_B}  \right)^2 - m_\pi^2   }\, .
\end{align}

Setting the lepton mass $m_\ell$ to zero one recovers the expression
\begin{align}  \label{eq:rate_nomass}
     \frac{\text{d} \Gamma(B \to \pi \, \ell \, \bar \nu_\ell) }{\text{d} q^2} & = \frac{8 \left| \vec p_\pi \right|  }{3} \frac{ G_F^2 \, \left| V_{ub} \right|^2 q^2 }{256 \, \pi^3 \, m_B^2}   \bigg[  H_0^2(q^2)  \bigg] \, ,
\end{align}
that holds to be an excellent approximation for $m_\ell = m_e$ or $m_\ell = m_\mu$. The helicity amplitudes $H_{0/t}$ are related to the form factors defined in Equation~\ref{eq:hadcurrent} as 
\begin{align}
 H_0 = \frac{2 m_B \, \left| \vec p_\pi \right| }{\sqrt{q^2}}\, f_+(q^2) \, , \\
 H_t = \frac{m_B^2 - m_\pi^2}{\sqrt{q^2}}\, f_0(q^2) \, . \label{eq:Ht}
\end{align}
 The form factors $f_{+/0}$ are functions of the four-momentum transfer squared and need to be calculated with non-perturbative methods. 
 
 Analyticity and unitarity impose strong constraints on heavy meson decay form factors~\cite{Grinstein:1992hq,Boyd:1994tt,
Arnesen:2005ez, Becher:2005bg, Bourrely:2008za}. In the following the form factors are parametrized using a series expansion what involves a mapping of the variable $q^2$ to the variable
\begin{align}
 z(t,t_0) = \frac{ \sqrt{t_+ - q^2} - \sqrt{ t_+ - t_0}   }{  \sqrt{t_+ - q^2} + \sqrt{ t_+ - t_0}  } \, ,
\end{align}
with
\begin{align}
t_{\pm}=\left( m_B \pm m_\pi \right)^2 , \,
t_0=\left( m_B + m_\pi \right)( \sqrt{m_B} - \sqrt{m_\pi} )^2  \, . \nonumber
\end{align}
In this expansion the form factors $f_+$ and $f_0$ are assumed to be analytic in $z$ expect for a branch cut in $[t_+, \infty)$ and poles in $[t_-,t_+]$. Using Blaschke factors these poles can be removed, c.f. Ref.~\cite{Boyd:1994tt,zexpansion}, and the form factors can be expanded as\footnote{This expression is recovered by starting with an expansion $P(z) \, \phi(z) \, f = \sum_n \, a_n \, z^n$ , with $P_i$ the Blaschke factor and by setting the outer functions $\phi$ to unity.} 
\begin{align} \label{eq:bcl}
 f_+(z) & =  \frac{1}{ 1 - q^2 / m_{B^*}^2} \sum_{n = 0}^{N_z - 1}\, b_j^+ \left[ z^n - (-1)^{n-N_z} \frac{n}{N_z} \, z^{N_z} \right] \, , \\
 \label{eq:series}
 f_0(z) & =  \sum_{n = 0}^{N_z - 1}\, b_j^0 \, z^n \, ,
\end{align}
with $m_B^* = 5.325$ GeV the $B^*$-meson mass, and $N_z$ the number of expansion parameters $b_j^{+/0}$.
Equation~\ref{eq:bcl} is known as the Bourrely-Caprini-Lellouch (BCL) expansion~\cite{Bourrely:2008za} in the literature. 

\begin{figure*}[t!]
\centerline{
\includegraphics[width=0.47\textwidth]{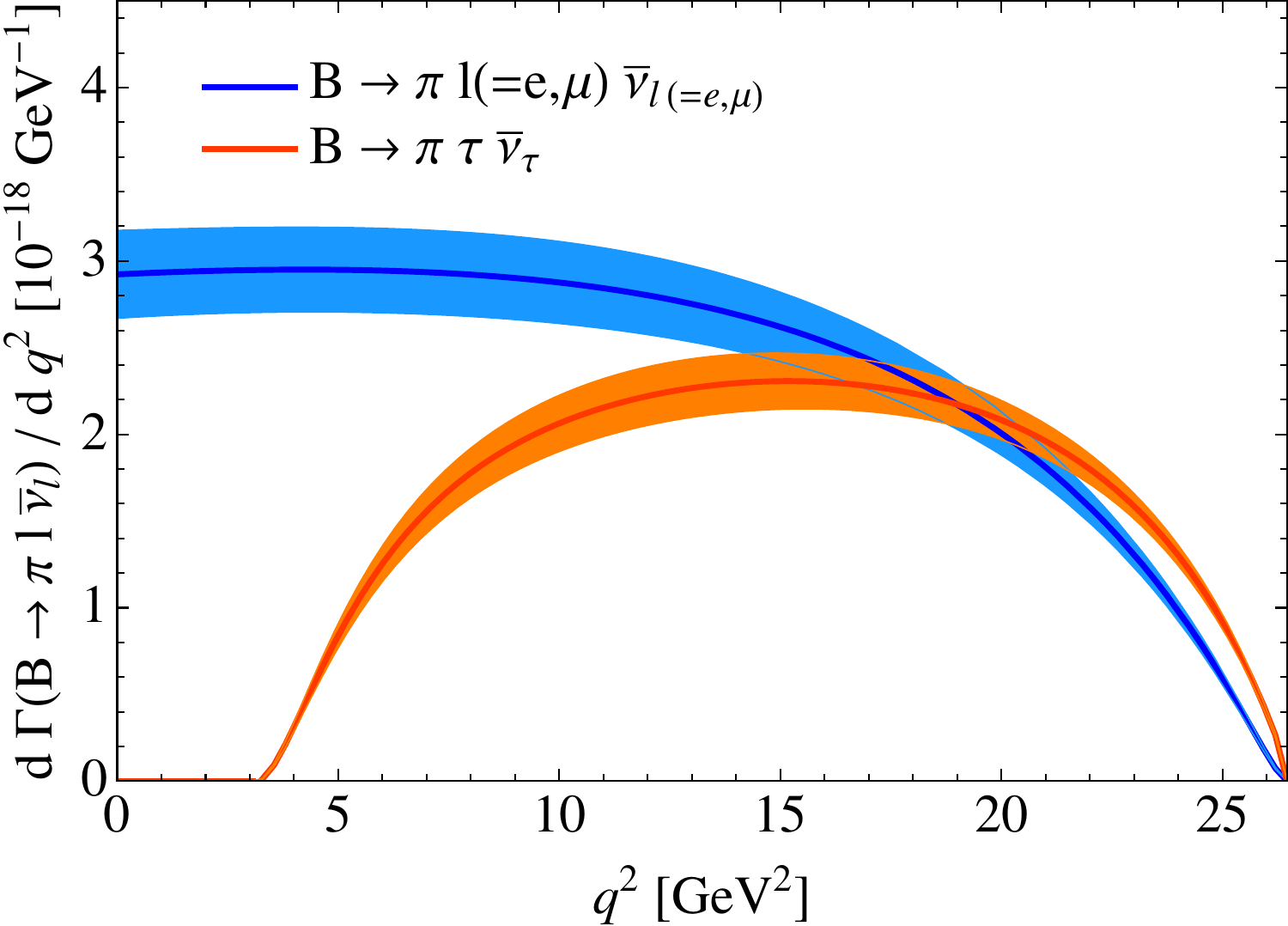} \hfill
\includegraphics[width=0.47\textwidth]{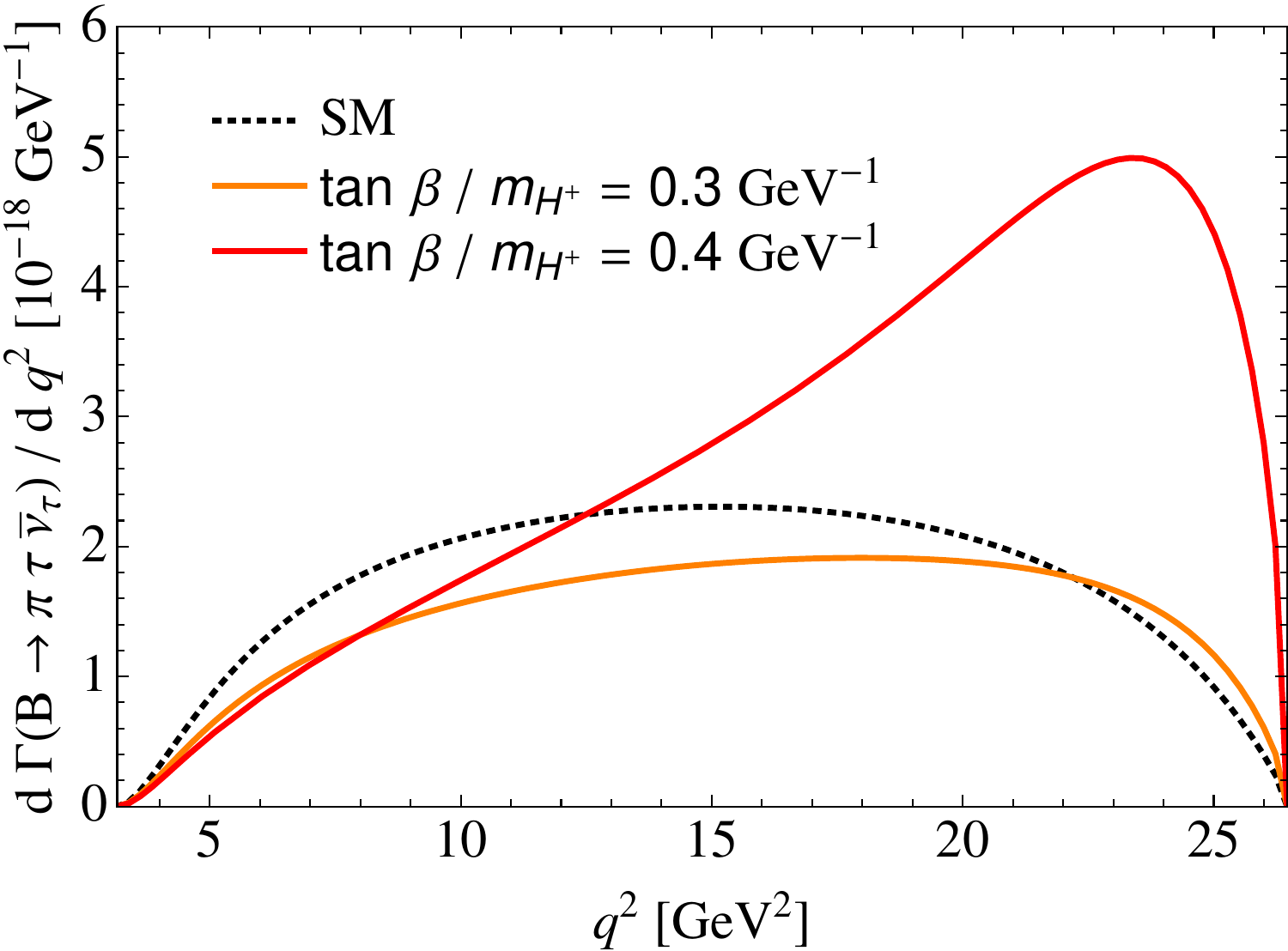}
}
\caption{ (left) The differential rate for $B \to \pi \, \ell \, \bar \nu_\ell$ with $\ell = e, \mu$ and  $B \to \pi \, \tau \, \bar \nu_\tau$ in the SM with associated errors are shown. As input for the form factors the 'Lattice+Exp' values of Table~\ref{tab:bcl_fermilabmilc} are used (right) The differential rate of $B \to \pi \, \tau \, \bar \nu_\tau$ is shown for the SM and two MSSM working points: $\tan \beta / m_{H^+} = 0.3$ ($H_t^{\rm 2HDM} < H_t^{\rm SM}$) and $\tan \beta / m_{H^+} = 0.4$ ($H_t^{\rm 2HDM} > H_t^{\rm SM}$).    }
\label{fig:bcl_mssm}
\end{figure*}

Ref.~\cite{FermilabMILC} recently reported (2+1)-flavor lattice QCD calculations of the two form factors $f_+$ and $f_0$ in the BCL expansion. In addition,  Ref.~\cite{FermilabMILC} also carried out a global analysis of the lattice QCD expansion parameters and all suitable $B \to \pi \, \ell \, \bar \nu_\ell$ partial branching fractions with the goal to extract $\left| V_{ub} \right|$. The post-fit expansion parameters from the combined analysis represent the current best knowledge of the $B \to \pi$ form factor, under the assumption that no new physics is affecting the light lepton decays. Both sets of expansion parameters are summarized in Table~\ref{tab:bcl_fermilabmilc} and Figure~\ref{fig:bcl_mssm} (left) shows the predicted differential rates for $B \to \pi \, \tau \, \bar \nu_\tau$ and $B \to \pi \, \ell \, \bar \nu_\ell$ using the expansion parameters from the global analysis.

\begin{table}[t!]
\begin{tabular}{cccccc}
\hline\hline
Lattice  & $b_0^i$ &   $b_1^i$ &  $b_2^i$ &  $b_3^i$ \\
\hline
$f_{i=+}$ & 0.407(15) & -0.65(16) & -0.5(9) & 0.4(1.3) \\
$f_{i=0}$ & 0.507(22) & -1.77(18) & 1.3(8) & 4(1)  \\
\hline
Lattice+Exp  &  $b_0^i$ &   $b_1^i$ &  $b_2^i$ &  $b_3^i$ \\
\hline
$f_{i=+}$ & 0.419(13) & -0.495(54) & -0.43(13) & 0.22(31) \\
$f_{i=0}$ & 0.510(19) & -1.700(82) & 1.53(19) & 4.52(83) \\
\hline\hline
\end{tabular}
\caption{The BCL expansion parameters of~\cite{FermilabMILC} are summarized: 'Lattice' shows the lattice only results and 'Lattice+Exp' the result of simultaneous fit the lattice information and all suitable measured partial branching fractions of $B \to \pi \, \ell \, \bar \nu_\ell$ with $\ell = e, \mu$. The full correlation information between the parameters can be found in Tables XIV and XIX of Ref.~\cite{FermilabMILC}. }
\label{tab:bcl_fermilabmilc}
\end{table}

\section{Modifications in the 2HDM type II}\label{sec:mssm}

In the two-Higgs doublet model extension of the SM, which describes the Higgs sector of the Minimal Supersymmetric model at tree level, the $B \to \pi \, \tau \, \bar \nu_\tau$ decay also can be mediated by a $H^\pm$ Higgs boson. Such an additional scalar mediator has drastic consequences as due to the Spin 1 nature of the $W^\pm$ boson the $B \to \pi$ transition occurs predominantly as a $P$-wave and is accordingly suppressed by a factor $\left| \vec p_\pi \right|^2$ near the maximal four-momentum transfer squared (where $\left| \vec p_\pi \right| \sim 0$). The $B \to \pi$ form factors reach their maximal value at the maximal four-momentum transfer and non-$P$-wave contributions from e.g. a scalar Higgs boson would greatly affect the rate in this part of phase space. 

The contributions of the charged Higgs Boson to $B \to \pi \, \tau \, \bar \nu_\tau$ decays can be incorporated into Equation~\ref{eq:rate} by the replacement~\cite{Tanaka:1994ay,kamenik}
\begin{align}\label{eq:2hdm}
H_t^{\rm SM}  \to H_t^{\rm 2HDM} \approx H_t^{\rm SM} \, \left( 1 - \frac{\tan^2 \beta}{m_{H^\pm}^2} \, \frac{q^2}{1 - m_u / m_b} \right) \, .
\end{align}
where $\tan \beta$ is the ratio of the vacuum expectation values of the two Higgs doublets and $m_{H^\pm}$ is the mass of the charged Higgs boson and $H_t^{\rm SM} = H_t$, cf. Eq.~\ref{eq:Ht}. Further $m_u / m_b$ is the ratio of the $u$- and $b$-quark masses at an arbitrary (but common) mass scale. This factor leads to a negligible modification and can in principle safely be ignored. Equation~\ref{eq:2hdm} proofs sufficiently accurate to probe charged Higgs bosons with a mass of more than about 15 GeV, what is the experimental interesting region as mass scales below 15 GeV are excluded by $b \to s \, \gamma$ measurements already~\cite{bsgamma} or other constraints~\cite{Crivellin:2013wna,Crivellin:2015mqa}.

Figure~\ref{fig:bcl_mssm} (right) compares the SM differential rate with two working points for $\tan \beta$ and $m_{H^+}$. The rate is only sensitive to their respective ratio and for a value of $\tan \beta / m_{H^+} = 0.3$ the negative interference causes $H_t^{\rm 2HDM}$ to be smaller than $H_t^{\rm SM}$, resulting in a lower $B \to \pi \, \tau \, \bar \nu_\tau$  rate than for the SM. For $\tan \beta / m_{H^+} = 0.4$ the additional scalar transition starts to dominate the matrix element, modifying the differential decay rate strongly near the kinematic endpoint. For larger values of $\tan \beta / m_{H^+}$ the 2HDM contribution completely dominates the total and differential rate.

\section{Prediction for $R_\pi$ and exclusion for $\tan \beta$ and $m_{H}^+$}\label{sec:rpi}

\begin{figure*}[t!]
\centerline{
\includegraphics[width=0.47\textwidth]{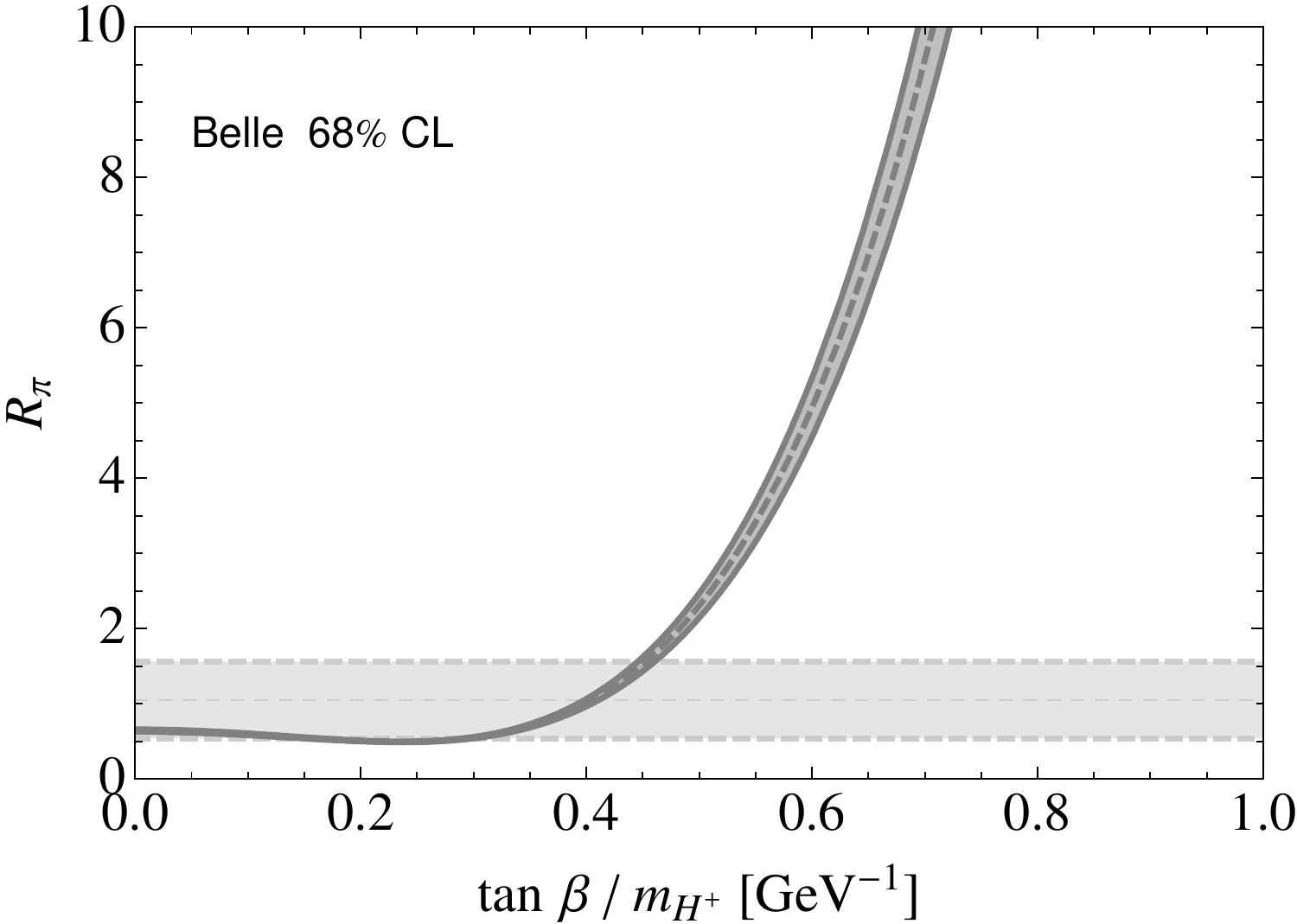}
\includegraphics[width=0.5\textwidth]{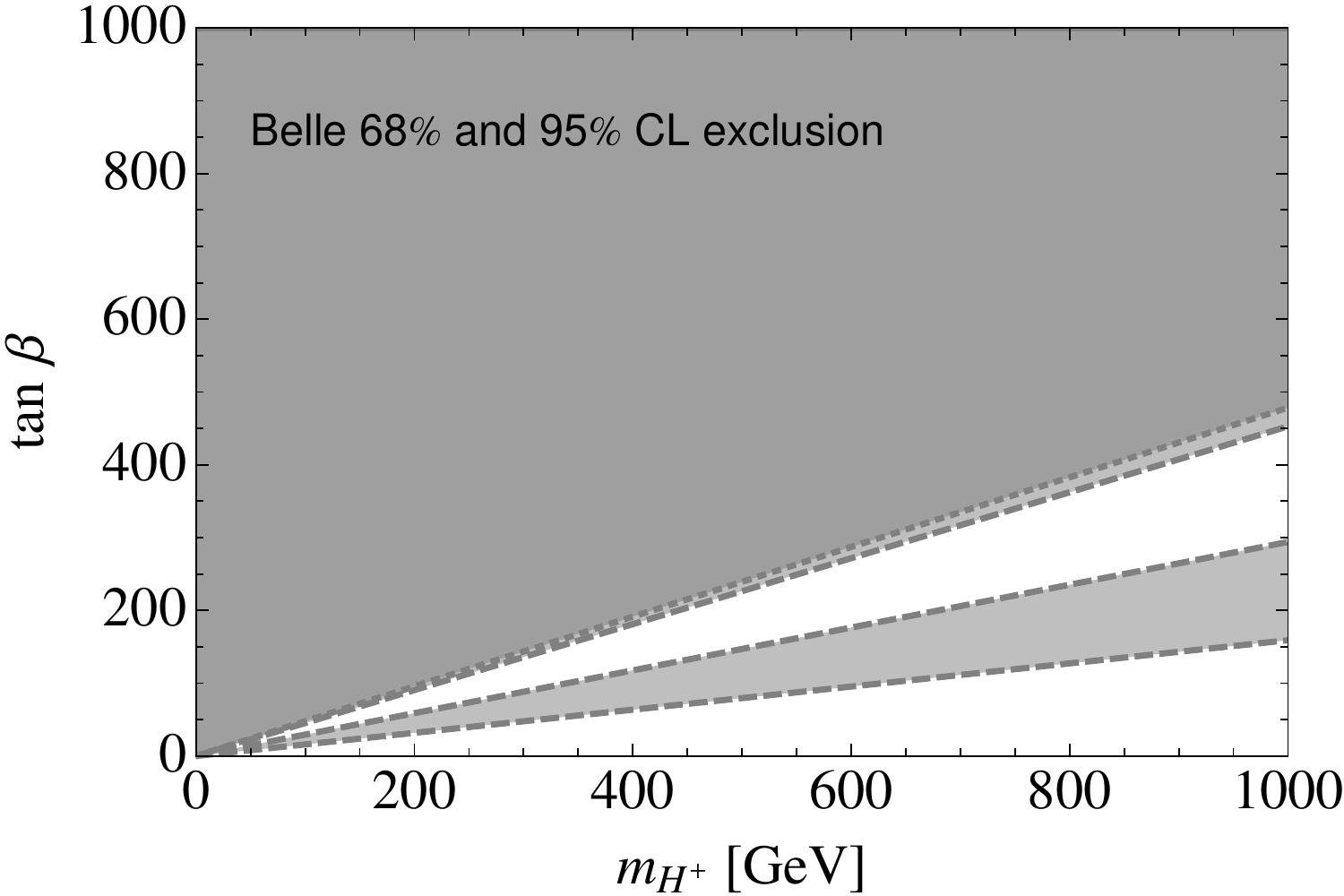}
}
\caption{ (left) The evolution of $R_\pi$ as a function of $\tan \beta / m_{H^+}$ is shown (dark grey). The experimental 68\% CI of Ref.~\cite{bellepitaunu} is shown (in light grey) and MSSM parameter points that predict ratios larger than $\approx 0.48$ can be excluded at 95 CL under the assumption that efficiencies and acceptances are modified negligibly with respect to the current statistical precision. (right) The 68\% (light grey) and 95\% (grey) excluded MSSM parameter points in the $\tan \beta-m_{H^+}$ plane.  }
\label{fig:rpi_tanb_mhp}
\end{figure*}

By integrating the differential rates Equations~\ref{eq:rate} and \ref{eq:rate_nomass} over the allowed kinematic range in $q^2$, a prediction for $R_\pi$ can be obtained:
\begin{align}
 R_\pi = \frac{ \Gamma(B \to \pi \, \tau \, \bar \nu_\tau) }{ \Gamma(B \to \pi \, \ell \, \bar \nu_\ell)  } = \frac{ \int_{m_\tau^2}^{q^2_{\rm max}}  \text{d}  q^2 \, \text{d}  \Gamma(B \to \pi \, \tau \, \bar \nu_\tau) / \text{d} q^2  }{  \int_{0}^{q^2_{\rm max}}  \text{d}  q^2 \, \text{d}  \Gamma(B \to \pi \, \ell \, \bar \nu_\ell) / \text{d} q^2 } \, ,
\end{align}
with $q^2_{\rm max}= m_B^2 + m_\pi^2 - 2 m_B m_\pi $ and the light lepton masses can be set to zero. The CKM matrix element $\left| V_{ub} \right|$, Fermi's constant, as well as other constant terms cancel in the ratio. Using the $B \to \pi$ BCL parameters from the global analysis of Table~\ref{tab:bcl_fermilabmilc} the following prediction for the SM value is obtained:
\begin{align}\label{eq:RpiSM}
 R_\pi^{\rm SM} = 0.641 \pm 0.016 \, .
\end{align}
The rates in the numerator and denominator are strongly correlated (correlation $\rho = 0.94$) if the integration is carried out over the full kinematic range, resulting in the large desired cancelation of non-perturbative uncertainties from the form factors. This number can be compared to the recently reported first measurement from Belle~\cite{bellepitaunu}:\footnote{The reported value for the branching fraction of Ref.~\cite{bellepitaunu} was divided by the current world average of the $B^0 \to \pi^- \, \ell^+ \, \nu_\ell$ branching fraction of $\left(1.45  \pm 0.05\right) \times 10^{-4}$ from Ref.~\cite{pdg}. }
\begin{align} \label{eq:meas}
 R_\pi^{\rm meas} = 1.05 \pm 0.51 \, .
\end{align}
Although statistically limited, this ratio already has some power to rule out several working points in $\tan \beta$ and $m_{H^+}$ as many predict a very strong enhancement of $R_\pi$. This can be seen in Figure~\ref{fig:rpi_tanb_mhp} (left) from the predicted (dark grey) and observed (light grey) $R_\pi$ values. 

To perform a proper exclusion in terms of $\tan \beta$ and $m_{H^+}$ though the impact on the acceptance and selection efficiencies should be studied. This can be done for instance by reweighing the used SM Monte Carlo (MC) samples to the probed MSSM working point with weights such as 
\begin{align}
 w & = \frac{ \Gamma_{\rm MC} }{ \Gamma(\tan \beta, m_{H^+}) } \frac{  \text{d} \Gamma(\tan \beta, m_{H^+}) / \text{d} q^2  }{  \text{d} \Gamma_{\rm MC} / \text{d} q^2  }
\end{align}
where the first factor ensures that the branching fraction in the simulation is unaltered and the quantities are $ \Gamma_{\rm MC}$ the total rate in MC, $\Gamma(\tan \beta, m_{H^+})$ the total rate at a given working point in $\tan \beta-m_{H^+}$, and $\text{d} \Gamma / \text{d} q^2$ the corresponding differential rates. In addition the fraction of negative and positive helicity $\tau$-leptons has to be corrected for as e.g. subsequent leptons are either emitted preferentially in or opposite to the $\tau$ direction. The evolution of the helicity fraction as a function of $\tan \beta / m_{H^+}$ is briefly discussed in appendix~\ref{sec:hel}.

Assuming that such effects are small with respect to the current statistical precision, a first exclusion can be carried out: Figure~\ref{fig:rpi_tanb_mhp} (right) shows the excluded region of $\tan \beta-m_{H^+}$ at 95\% CL (dark grey) and 68\% CL (light grey). A comparison with the current bounds from $B \to \tau \, \bar \nu_\tau$ can be found in appendix~\ref{sec:btaunu}.

\section{Other observables with increased sensitivity}\label{sec:rpiq2}

An interesting alternative observable besides $R_\pi$ could be the ratio of partially integrated and measured rates with a lower cut on $q^2$. This is motivated that in the 2HDM type II for large values of $\tan \beta / m_{H^+}$ the constructive interference is dominating the rate Equation~\ref{eq:rate} and due to the absence of the $P$-Wave suppression strongly modifies the large $q^2$ region. A fully differential analysis of course contains the most power to probe and distinguish such a scenario with a scalar mediator from the SM, but with the current small experimental sensitivity a measurement that probes the high $q^2$ range in one bin already could lead to an improved rejection of large $\tan \beta / m_{H^+}$ ratios. For instance: the SM prediction for a measurement in $[q^2_{\rm max} / 2 , q^2_{\rm max}]$ is
\begin{widetext}
\begin{align}\label{eq:RpiSM_diff}
 R_\pi^{\rm SM} (q^2_{\rm min} = q^2_{\rm max}   / 2) = \frac{ \int_{q^2_{\rm min}= q^2_{\rm max}   / 2}^{q^2_{\rm max}}  \text{d}  q^2 \, \text{d}  \Gamma(B \to \pi \, \tau \, \bar \nu_\tau) / \text{d} q^2  }{  \int_{q^2_{\rm min}= q^2_{\rm max}   / 2}^{q^2_{\rm max}}  \text{d}  q^2 \, \text{d}  \Gamma(B \to \pi \, \ell \, \bar \nu_\ell) / \text{d} q^2 } = 1.012 \pm 0.008 \, .
\end{align}
\end{widetext}
and the ratio of branching fractions at for example $\tan \beta / m_{H^+} = 0.4$ results in a prediction about a factor of two larger for $ R_\pi(q^2_{\rm max} / 2) = 2.09 \pm 0.03$. In comparison, the fully inclusive rate ratio for the same $\tan \beta / m_{H^+}$ working point results in $R_\pi = 1.01 \pm 0.04$ and shows a smaller enhancement with respect to the SM value Eq.~\ref{eq:RpiSM}. The reduction in experimental sensitivity on $R_\pi$ due to only analyzing half of the allowed phase space of course also has to be taken into account: a rough estimate for the increase in statistical error, assuming that background is distributed uniformly in $q^2$ and that the signal is SM like, is a factor of about $\approx \sqrt{2}$. These numbers imply an overall improved sensitivity for a measurement in $[q^2_{\rm max} / 2 , q^2_{\rm max}]$ over the fully inclusive $R_\pi$ value to exclude $\tan \beta / m_{H^+}$ working points. In addition, as pointed out already by Refs.~\cite{Freytsis:2015qca,Tanaka:1994ay}, the cancellation of the non-perturbative error further improves allowing for a more precise predictions of the ratio of partial rates.

\section{Summary and Conclusion}\label{sec:concl}

The $B \to \pi \, \tau \, \bar \nu_\tau$ decay offers an interesting alternative decay to probe the deviations observed in semi-tauonic decays with charmed final states. In this manuscript a prediction for the ratio of semi-tauonic and light lepton total rates, $R_\pi$, is presented using the state-of-the-art knowledge of the $B \to \pi$ form factor. In the SM a value of $R_\pi = 0.641 \pm 0.016$ is found. The impact of the presence of charged Higgs boson contributions in the context of the 2HDM type II is discussed and a first preliminary exclusion in $\tan \beta - m_{H^+}$ using a recent Belle measurement is carried out. In addition the idea is presented to measure in future measurements the high $q^2$-region, as it has a higher sensitivity to probe the MSSM parameter space in the 2HDM type II model. 

%\vspace{4ex}

\section*{Acknowledgements}

%\vspace{-4ex}

The author expresses thanks to Bob Kowalewski, Christoph Schwanda, Stephan Duell and Jochen Dingfelder for their useful comments that helped improve the manuscript. A special thanks also goes to Elena and Savanna Yu for good conversations in Copenhagen that played an important role in the creation of this manuscript.

\clearpage

\begin{appendix}

\begin{figure*}[p!]
\centerline{
\includegraphics[width=0.48\textwidth]{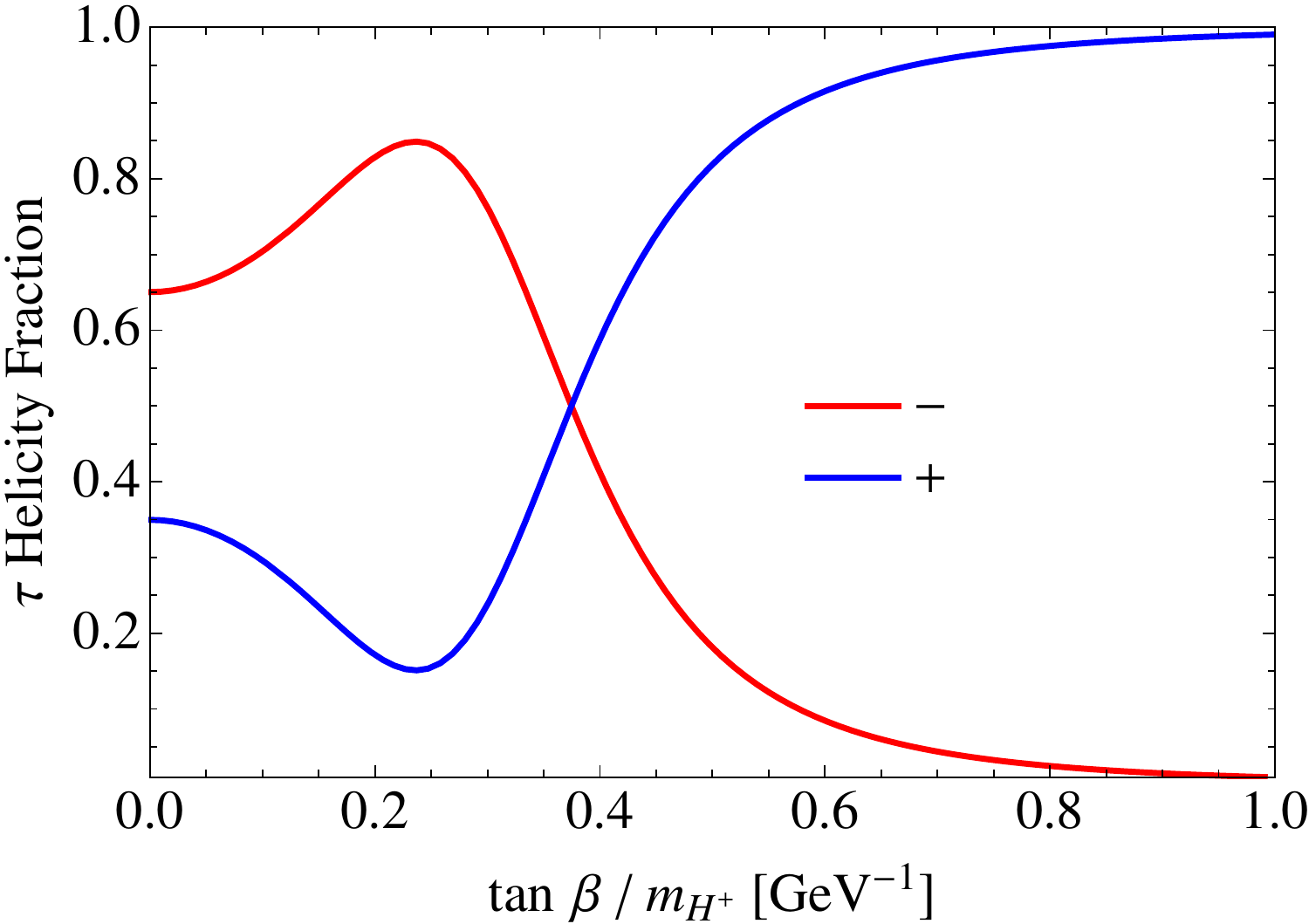}
\includegraphics[width=0.50\textwidth]{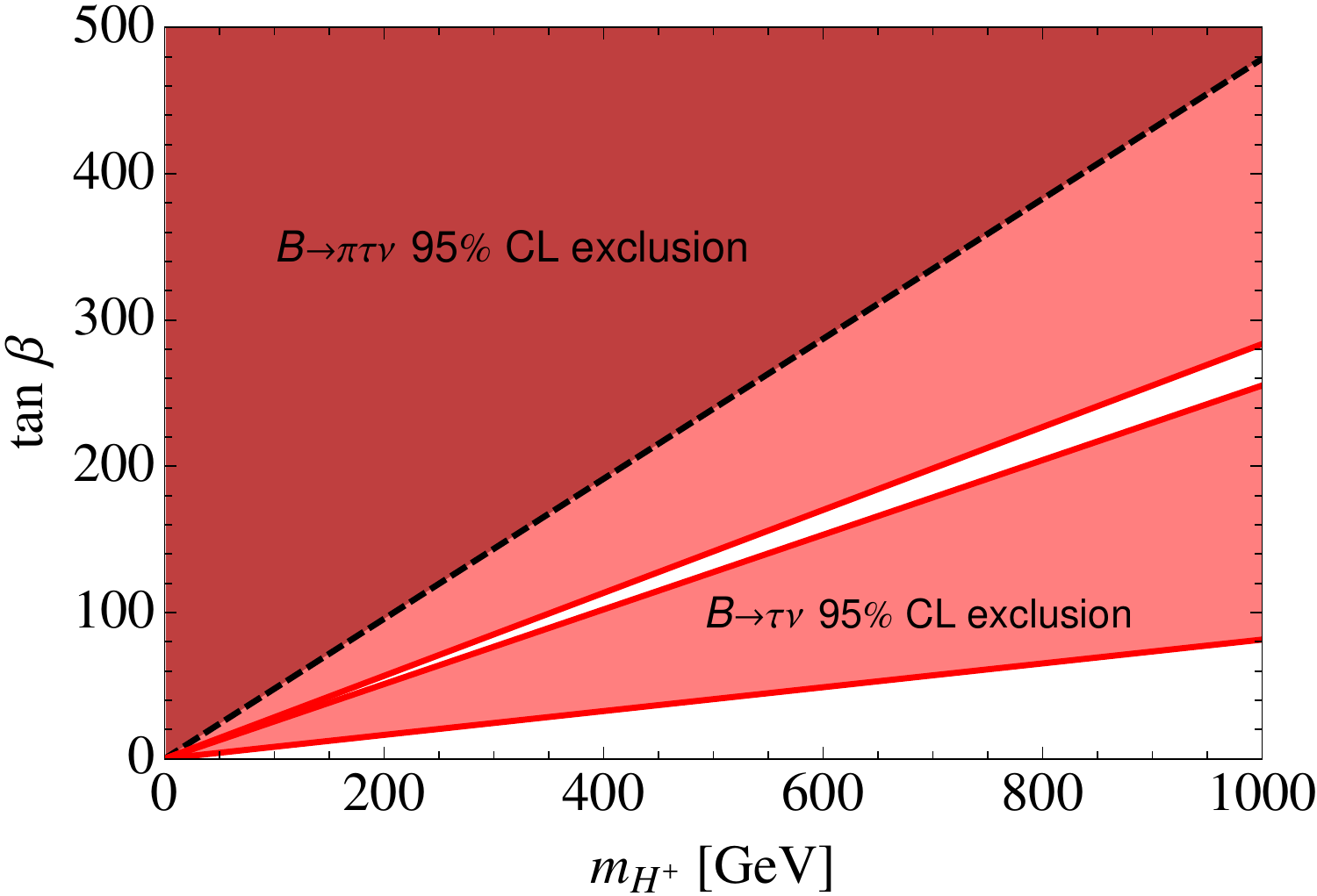}
}
\caption{ (left) The $\tau^-$-lepton positive (+) and negative (-) helicity fractions as a function of $\tan \beta / m_{H^+}$ for $B \to \pi \, \tau^- \, \bar \nu_\tau$ is shown. (right) The 95\% exclusion limit from Ref.~\cite{bellepitaunu} is compared with the exclusion limits from $B \to \tau \, \bar \nu_\tau$. As input an averaged branching fraction of $\mathcal{B}(B \to \tau \, \bar \nu_\tau) = \left( 1.03 \pm 0.20 \right) \times 10^{-4}$ and $\left| V_{ub} \right| = \left( 3.72 \pm 0.16 \right) \times 10^{-3}$ was used. 
%(right) The  $\tau$-lepton helicity fraction as a function of $q^2$ for the SM and  $\tan \beta / m_{H^+} = 0.5$.   
}
\label{fig:tau_hel}
\end{figure*}

\section{Helicity fractions in the 2HDM type II}\label{sec:hel}

In the SM the $\tau^-$ polarization is about 35\% positive and 65\% negative. This relative fraction changes dramatically in the 2HDM type II: in the range of $\tan \beta / m_{H^+} \in [0, 0.25]$ the negative helicity fraction increases to almost 85\%. This has drastic consequences if the $\tau^-$-leptons are reconstructed using e.g  leptonic channels, $\tau^- \to \ell^- \, \bar \nu_\tau \, \bar \nu_\ell$: the secondary lepton spectrum in the $B$-meson rest frame will become harder with respect to the SM spectrum as the lepton is preferably emitted in the direction of the $\tau^-$-lepton altering the acceptance and selection efficiencies. After this turning point the positive helicity starts to dominate and at $\tan \beta / m_{H^+} = 0.7$ about 95\% of the $\tau^-$-leptons have positive helicity. If reconstructed again via $\tau^- \to \ell^- \, \bar \nu_\tau \,\bar \nu_\ell$, the secondary lepton spectrum now is softer than the SM spectrum as the leptons preferably are emitted opposite to the $\tau^-$-lepton direction. Figure~\ref{fig:tau_hel} (left) shows the helicity fractions as a function of $\tan \beta / m_{H^+}$ for $B \to \pi \, \tau^- \, \bar \nu_\tau$.
%Figure~\ref{fig:tau_hel} (right) shows the evolution of the $\tau$ helicity fractions as a function of $q^2$ for the SM and  $\tan \beta / m_{H^+} = 0.4$.

\section{Comparison with $B \to \tau \, \bar \nu_\tau$}\label{sec:btaunu}

Figure~\ref{fig:tau_hel} (right) compares the 95\% exclusion limits of the 2HDM type II parameter space from  $B \to \pi \,  \tau \, \bar \nu_\tau$ with the exclusion from $B \to \tau \, \bar \nu_\tau$: Both decays involve the same quark lines and the existing measurements for $B \to \tau \, \bar \nu_\tau$ result in tighter constraints as the individual $B \to \pi \,  \tau \, \bar \nu_\tau$ branching fraction. The two red regions is from $B \to \tau \, \bar \nu_\tau$ and the dark red region (dashed black boundary) is from $B \to \pi \,  \tau \, \bar \nu_\tau$.  As input a world average of $\mathcal{B}(B \to \tau \, \bar \nu_\tau) = \left( 1.03 \pm 0.20 \right) \times 10^{-4}$ was used, obtained from averaging the latest Belle results~\cite{Adachi:2012mm,Kronenbitter:2015kls} with the values reported from \babar~\cite{Lees:2012ju}. Further, for $\left| V_{ub}\right|$ and the $B$-meson decay constant the reported values of Ref.~\cite{FermilabMILC} and \cite{Christ:2014uea} were used, respectively.

\end{appendix}

\end{document}